# ON THE FORMATION OF MULTIPLE ABSORPTION TROUGHS IN BROAD ABSORPTION LINE QSOS

Nicolas A. Pereyra[*]
pereyrana@utpa.edu
Department of Physics and Geology, University of Texas – Pan American

**Abstract**

We present theoretical C IV $\lambda\lambda$1548,1550 absorption line profiles for QSOs calculated assuming the accretion disk wind (ADW) scenario. The results suggest that the multiple absorption troughs seen in many QSOs may be due to the discontinuities in the ion balance of the wind (caused by X-rays), rather than discontinuities in the density/velocity structure. The profiles are calculated from a 2.5D time-dependent hydrodynamic simulation of a line-driven disk wind for a typical QSO black hole mass, a typical QSO luminosity, and for a standard Shakura-Sunyaev disk. We include the effects of ionizing X-rays originating from within the inner disk radius by assuming that the wind is shielded from the X-rays from a certain viewing angle up to $90^o$ (``edge on"). In the shielded region we assume constant ionization equilibrium, and thus constant line-force parameters. In the non-shielded region we assume that both the line-force and the $C^{+3}$ populations are nonexistent. The model, at viewing angles close to the angle that separates the shielded region and the non-shielded region, produces absorption lines with multiple troughs. The steady nature of accretion disk winds, in turn, may account for the steady nature of the absorption structure observed in multiple-trough broad absorption line QSOs. The model parameters are $M_{bh} = 10^9\,M_\odot$ and $L_{disk} = 10^{47}$ erg s$^{-1}$.

[*] Corresponding author: Fax 1 956 665 2423
 E-mail address: pereyrana@utpa.edu  (Nicolas A. Pereyra)

# LA FORMACION DE MULTIPLES DEPRESIONES DE ABSORPTION EN QUASARES DE LINEA DE ABSROPTION ANCHA

Nicolas A. Pereyra[*]
pereyrana@utpa.edu
Department of Physics and Geology, University of Texas – Pan American

**Abstract**

Presentamos aquí líneas espectrales teóricas de absorción de C IV $\lambda\lambda$1548,1550 para quásares (QSOs) calculados asumiendo el modelo de vientos de discos de acreción (ADW). Los resultados sugieren que las múltiples líneas-anchas de absorción observado in muchos quásares puede ser causado por una discontinuidad en el equilibrio iónico del viendo (debido a presencia de rayos-X), y no debido a la discontinuidad de densidad y/o velocidad del viento. Las líneas espectrales son calculadas de una simulación hidrodinámica 2.5D de un viento de línea de disco usando valores típicos de masa del hoyo negro de un quásar, un luminosidad típica de quásar, y un disco estándar de Shakura-Sunyaev. Aquí incluimos los efectos de rayos-X ionizante originándose dentro del radio interno del disco a través de la suposición de que los rayos-X son bloqueados de llegar al viento desde un ángulo dado hasta 90º (disco visto "de lado"). En la región en donde están bloqueada los rayos-X, asumimos un equilibrio iónico local constante y por lo tanto parámetros de fuerzas-de-línea constantes. En la región donde los rayos-X no están bloqueados, asumimos que tanto la fuerza-de línea y las poblaciones de $C^{+3}$ son inexistentes. El modelo, en ángulos cerca del ángulo que separa la región donde están bloqueado los rayos-X y la región donde no están bloqueada los rayos-X, se producen múltiples líneas-anchas de absorción. El carácter estable de vientos de discos de acreción podrían, a su vez, podría explicar la naturaleza estable de la estructura de absorción observadas en quásares que presentan múltiples líneas anchas de absorción. Los parámetros del modelo son $M_{hn} = 10^9 \, M_\odot$ y $L_{disco} = 10^{47}$ erg s$^{-1}$.

*Palabras Claves:* Acreción, Discos de Acreción, Hidrodinámica, Quásares, Líneas de Absorción

---

[*] Autor de contacto: Fax 1 956 665 2423
Correo Electrónico: pereyrana@utpa.edu  (Nicolas A. Pereyra)

# 1. Introduction

Approximately 10% of QSOs are observed to present Broad Absorption Lines (BALs) (e.g., Turnshek [43]). Two recent studies illustrate the two scenarios that have been invoked to explain the detection of BALs in only a fraction of QSOs. Boroson [4] suggests that the differences in QSO spectral features may primarily be due to QSOs with intrinsically different properties. On the other hand, Elvis [11] stresses the importance of viewing angle in modulating several observed spectral features of QSOs. Both views present some level of truth. Surely not all QSOs have the same intrinsic properties such as black hole mass; however if the widely held view that accretion disks power QSOs/AGN is valid (e.g., Lin et al. [25]; Ulrich et al. [45]; Mirabel et al. [28]) some viewing angle effect must be present.

BAL QSOs exhibit a wide variety of profile shapes: classic P-Cygni (i.e., a blue shifted absorption trough superimposed with an emission line) (Lynds [26]; Burbidge [7]; Burbidge [8]; Scargle et al. [36]; Turnshek et al. [44]); detached absorption troughs (Osmer et al. [31]), and multiple absorption troughs (Turnshek et al. [40]).

Murray et al. [29], using one-dimensional streamline models, showed that with an appropriate X-ray shielding mechanism, an accretion disk wind (ADW) driven by line radiation pressure with streamlines approximately parallel to the disk at high velocities could roughly account for: the flow velocity inferred in BAL QSOs, the detached troughs observed in many QSOs, and the fraction of QSOs with BALs. In the work of Murray et al. [29] the detection of BALs is dependent on viewing angle. The work we present in this paper is an extension of the work by Murray et al. [29] in that we continue to explore the ADW scenario for QSOs.

X-ray luminosities inferred in QSOs are typically comparable with their corresponding UV/optical luminosities (e.g., Tananbaum et al. [39]; Grindlay et al. [19]; George et al. [16]). This is consistent with the X-ray emission being powered by the accreting mass within the inner disk radius. BAL QSOs are observed to be X-ray quiet (Kopko et al. [23]; Green et al. [17]; Gallagher et al. [13]). Evidence that this lack of X-rays is due to absorption of an otherwise normal QSO X-ray source has been presented based on analysis of ROSAT, ASCA, and Chandra X-ray observations (Brandt et al. [5]; Mathur et al. [27]; Green et al. [18]; Gallagher et al. [14]). This is consistent with the idea that BALs are observed close to "edge on," with the line-of-sights falling within the X-ray shielded region. X-ray variability, with timescales of less than one day, is common in Active Galactic Nuclei (Mushotzky et al. [30]), and in particular, has been observed in QSOs (Hayashida et al. [21]). These timescales imply upper limits (Terrel [39]) on the size of the X-ray emitting region which are consistent with the X-ray emission occurring within the inner disk radius.

The BAL profiles exhibiting multiple absorption troughs have been generally interpreted as a consequence of discrete density/velocity flow distributions (Turnshek et al. [40]; Turnshek [41]; Turnshek [42]; Foltz et al. [12]; Braun et al. [6]; Korista et al. [24]; Arav et al. [2]; de Kool et al. [9]). Several possible mechanisms to account for a discrete character of the flow have been suggested: discrete ejection episodes of flow material (Turnshek et al. [41]), line-locking (Foltz et al. [12]; Korista et al. [24]), and separate outflows (Arav et al. [2]).

We show here that a more likely explanation is that the multiple absorption troughs are caused by discontinuities in the wind ionization balance, due to the presence of X-rays, within a steady disk wind flow. In our models, the presence of X-rays originating at the disk's center, generates two regions: a shielded region, extending from the disk surface down to a specific viewing angle, where we assume constant ion populations; and a non-shielded region where the radiative line-force and the $C^{+3}$ ionization fractions are nonexistent.

The steady nature of accretion disk winds (Pereyra et al. [34]) can account for the steady nature of the absorption structure which is observed in multiple (and in detached and P-Cygni) absorption troughs (Foltz et al. [12]; Smith et al. [38]; Turnshek et al. [45]; Barlow et al. [3]; Hamann et al. [20]). For typical QSO parameters, we develop 2.5D time-dependent hydrodynamic models of line-driven disk winds. From these models we calculate theoretical C IV line profiles, under the assumption of single scattering, and find that the multiple absorption troughs appear when the viewing angle is near the angle separating the shielded and the non-shielded regions.

In section §2 we describe this model. Results are presented in §3, and a summary is presented in §4.

## 2. Model

Our 2.5D time-dependent hydrodynamic ADW models (Hillier et al. [22]) extend the work of Murray et al. [29], who studied the ADW scenario through 1D streamline models. The ADW scenario (Figure 1) is a unified model of BAL and non-BAL QSOs in which the BALs are formed in an accretion disk wind driven by line radiation pressure. Within the ADW scenario the BALs are detected when the accretion disk is observed at high inclination angles (close to "edge on").

The 2.5D time-dependent hydrodynamic models are based on the PPM numerical scheme (Collela et al. [10]). The model is solved in cylindrical coordinates under azimuthal symmetry, similar to the line-driven accretion disk wind models developed by Pereyra [32] and Pereyra et al. [33] for cataclysmic variables. Within the standard accretion disk model (Shakura et al. [37]), the accretion disk accounts for the observed UV/optical continuum emission.

In these exploratory calculations we have assumed, for simplicity, that the X-rays are emitted at disk center and are shielded by the disk itself. Future modeling will explore other shielding mechanisms, including shielding by a failed inner disk wind (or hitchhiking gas). In the X-ray shielded region we assume constant ion populations and implement constant line-force parameters for the line radiation force (see Abbott [1] and Galey [15] for discussions on these parameters). In the non-shielded region the line-force is assumed to be nonexistent.

We use the Sobolev approximation (Rybicki et al. [35]) to compute the line-force. The full angular dependence of the radiation field coming from the disk is taken into account. Further, we allow for velocity gradients in both vertical and radial directions. The disk material is rotating about the black hole, the radial gravitational forces are at balance with the centrifugal forces, and the wind must start off vertically. As the material lifts off the disk it is increasingly exposed to the strong UV radiation field originating from inner disk radii. As a consequence the additional acceleration is primarily in the radial direction, and the wind then tends to flow radially.

Due to the presence of X-rays, and in particular with more realistic X-ray models, it is unlikely that a radiation wind can be driven at all radii. Our models therefore assume a minimum cutoff radii, which for the models presented here is taken as 20 $r_0$, where $r_0 = 6GM/c^2$ is the inner radius of the standard Shakura-Sunyaev disk.

As noted above, the earlier models of Murray et al. [29] were able to account for both the P-Cygni absorption troughs and the detached absorption troughs in QSOs; in their models it was assumed that the whole disk-wind-flow region was soft-X-ray shielded. The changes in ionization balance considered by Murray et al. [29] were due to decreasing densities along the streamlines which resulted in smooth detached and P-Cygni absorption troughs with deeper absorption at lower velocities.

In the models we present here, some of the wind, although necessarily starting in the X-ray shielded region, can flow towards the non-shielded region. When the wind crosses into the non-shielded region there is a dramatic change in the wind ionization. As discussed below, this may allow for the creation of multiple absorption troughs at some viewing-angles.

## 3. Results

The results presented here are for black hole mass $M_{bh} = 10^9$ $M_\odot$ and disk luminosity $L_{disk} = 10^{47}$ erg s$^{-1}$. The associated mass accretion rate within the standard disk model (Shakura et al. [37]) is $dM_{accr}/dt = 21$ $M_\odot$ yr$^{-1}$; the corresponding Eddington ratio $\sigma L_{disk} / 4\pi cGM_{bh} = 0.80$ .

The X-rays are assumed to be shielded between 90$^o$ ("edge on") and 80$^o$. Depending on the specific spatial distribution of X-ray emission within the inner disk radius, and assuming the vertical height distribution of a standard accretion disk, for the above physical parameters the disk itself could account for X-ray shielding from ~71$^o$ to 90$^o$.

The line-force parameters used are: k = 0.002 and $\alpha$ = 0.6 (Abbott [1] and Murray et al. [29]). Because the QSO radiation is substantially harder than for O stars, the wind is more ionized and k is substantially lower. For the above parameters we find a wind mass loss rate of dMwind/dt ≈ 10$^{-2}$ $M_\odot$ yr$^{-1}$, which is much less than the accretion rate. In our current model the origin of the line-driving radiation is distributed over a disk surface rather than a point source as in the earlier models of Murray et al. [29]; as a consequence of this, we are finding lower wind mass loss rates than the earlier 1D streamline models of Murray et al. [29].

Using the parameters described above, or similar parameters, we find steady disk winds from our models (Figure 2). We confirm earlier findings by Murray et al. [29] that the ADW scenario can account for the form of both the P-Cygni type and the detached type absorption troughs. Further, we find that multiple absorption troughs could also be accounted for.

In Figure 3 we present theoretical C IV absorption line profiles.The line profiles are calculated by integrating over the disk, consistent with the calculation of the driving radiation flux in the hydrodynamic calculations. The strength and complexity of the multiple absorption troughs vary among the different QSOs in which they are observed (Turnshek [42]). However, one can find BAL QSOs that present absorption structures in C IV λλ1548,1550 similar to the theoretical multiple absorption trough profile we present in Figure 3 (at a viewing angle of 83º). For example, see Q0335-336 (Turnshek [43]), and Q0226-1024 (Korista et al. [24]), and Figure 4.

In our models the multiple absorption troughs are generated by discontinuities in the ionization balance of the wind; these discontinuities in turn are caused by the presence of X-rays (see Figure 5), the disk wind in itself presenting a smooth steady flow. The multiple troughs are observed at viewing angles close to the separation between the X-ray shielded and non-shielded regions, the P-Cygni troughs are observed closer to "edge on", and the detached troughs are observed at angles in between. This model accounts for the steady absorption structure which is observed in multiple (and in detached and P-Cygni) absorption troughs (Foltz et al. [12]; Smith et al. [38]; Turnshek et al. [44]; Barlow et al. [3]; Hamann et al. [20]).

The formation of steady multiple troughs in our models are a consequence of the steady nature of the wind and of the coupling between the X-rays, wind ionization balance, and hydrodynamics. As we discussed in the Introduction, the X-ray luminosities and their variability are consistent with the X-ray emission powered by accreting mass within the inner disk region; thus the X-ray emission region will be much smaller than the disk itself, and in our current models it is treated as a point source at disk center which is shielded at a given angle. The X-rays will ionize the wind; in our current work we are modeling this effect by assuming nonexistent line-force and nonexistent $C^{+3}$ populations in the non-shielded region. Although our models will evolve towards more realistic ones, the idea of X-ray ionization effects influencing BAL profiles and the trough structure is important.

## 4. Summary

We are attempting to explore the ADW scenario in greater detail than has been previously done, with the goal of studying the potential and limitations of the ADW scenario to describe the general properties of QSO spectra. Presented here are some of our initial efforts in which we extend the 1D streamline models of Murray et al. [29], to 2.5D. Within the parameters we have explored we find steady disk winds, and confirm that disk winds can potentially account for BALs observed in some QSOs.

It is found that the multiple absorption troughs, observed in many BAL QSOs, could be accounted for by discontinuity in the ionization balance of the outflow, rather than by discontinuity in the density/velocity structure of the outflow as has been suggested previously by many different authors.

In turn, the steady nature of ADWs may account for the steady nature of the velocity structure of absorbing gas, that has been inferred for BAL QSOs through the observation of the C IV λλ1548,1550 absorption line profile at different epoch.

We are currently developing more detailed 2.5D models as well performing more detailed analysis of our models. In particular, with respect to the formation of steady multiple troughs, we will analyze the steady nature of the wind and its coupling with the ionizing X-rays. As we have shown here, within the ADW scenario, these two characteristics may be important in the formation of multiple absorption troughs in QSOs.

**Figure Captions**

Fig. 1 Accretion Disk Wind Model for QSOs.

Fig 2. Results from one of our calculated models. Vector field graph of wind velocity superimposed with electron density contours. Contour levels vary by factors of two from $1.5 \times 10^5$ cm$^{-3}$ down to $0.006 \times 10^5$ cm$^{-3}$. The lower density wind values are at the top of the graph and at the lower right corner. The black hole is at the origin and the disk is along the horizontal axis. The straight thick line indicates the border between the X-ray ionized region (above) and the X-ray shielded region (below).

Fig. 3. Theoretical C IV $\lambda\lambda$1548,1550 line profiles, obtained from the model of Figure 2, assuming single scattering, and that $C^{+3}$ is the dominant C ionization stage. The line profiles shown by the dotted lines are calculated under the assumption that radiation is absorbed only. The viewing angle is indicated above each graph, as measured with respect to the disk rotation axis ($0^o$ corresponding to face-on). For the model of Figure 2, our line profile calculations indicate that most of the line formation is occurring between $r = 20$ $r_0$ and $r = 50$ $r_0$. Note the formation of multiple absorption troughs at $83^o$.

Fig. 4 Observational spectra and C IV $\lambda\lambda$1548,1550 line profiles for two BAL QSOs presenting multiple absorption troughs; taken from the Sloan Digital Sky Survey database: SDSS J005355.15-000309.31 (left) and SDSS J011227.6-011221.76 (right). The dashed line is the continuum spectra.

Fig 5. Theoretical C IV $\lambda\lambda$1548,1550 line profiles, obtained from the model of Figure 2, assuming pure absorption at a viewing angle of $83^o$. The dotted line is the same as in Figure 3. The dashed line is calculated under the assumption of constant ionization equilibrium throughout (including the non-shielded region). This figure illustrates that the formation of multiple absorption troughs, within our models, are a consequence of the discrete character of the ionization balance in a steady flow.

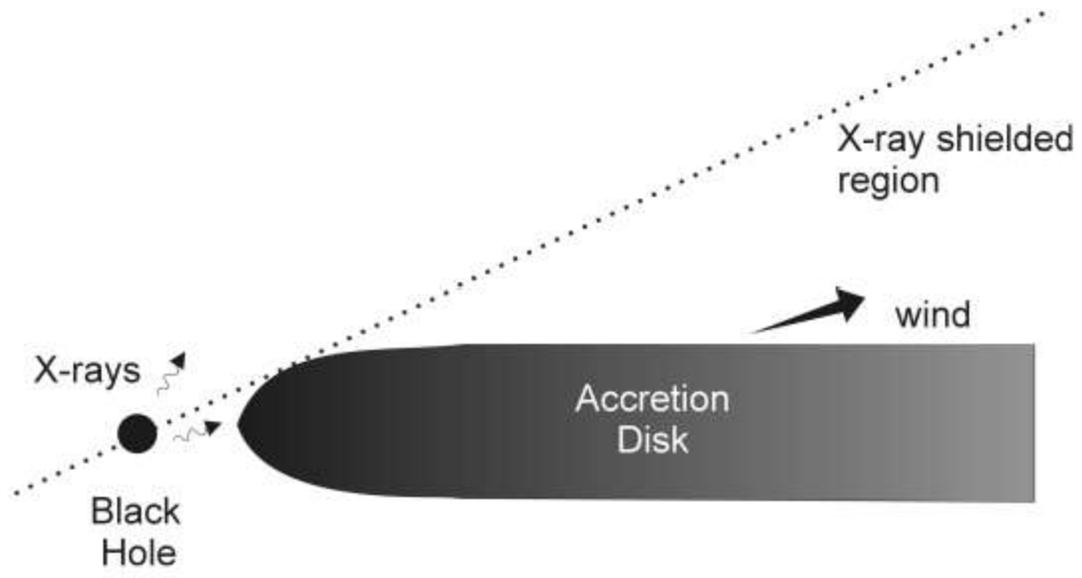

Fig. 1

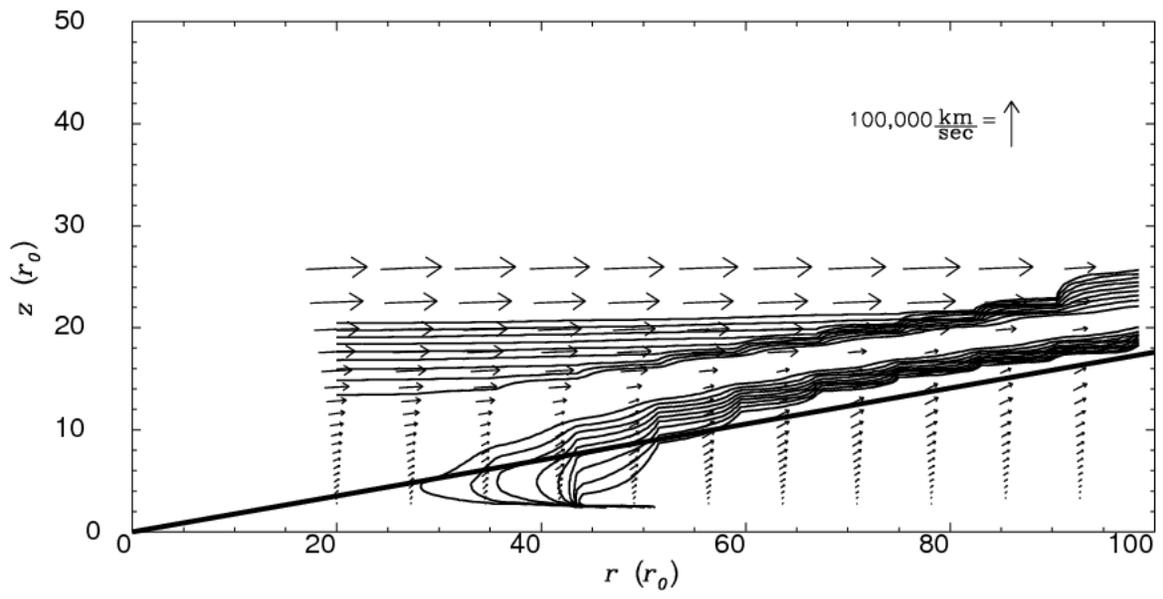

Fig. 2

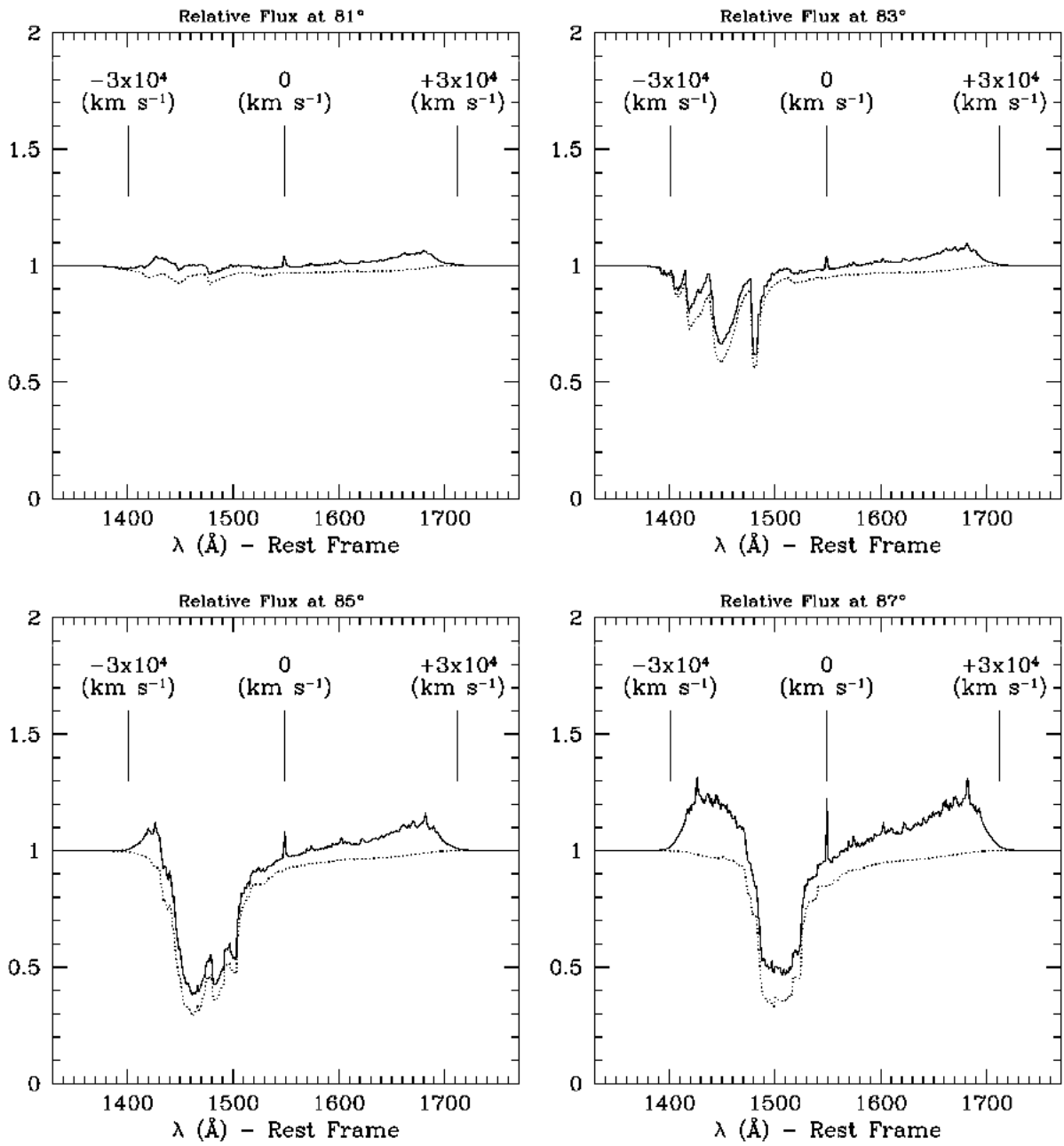

Fig. 3

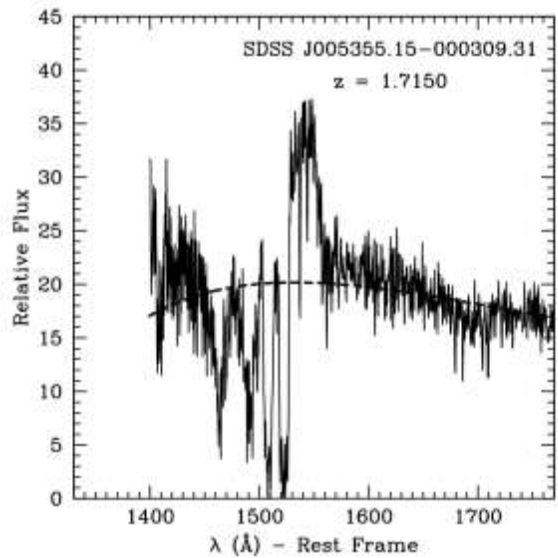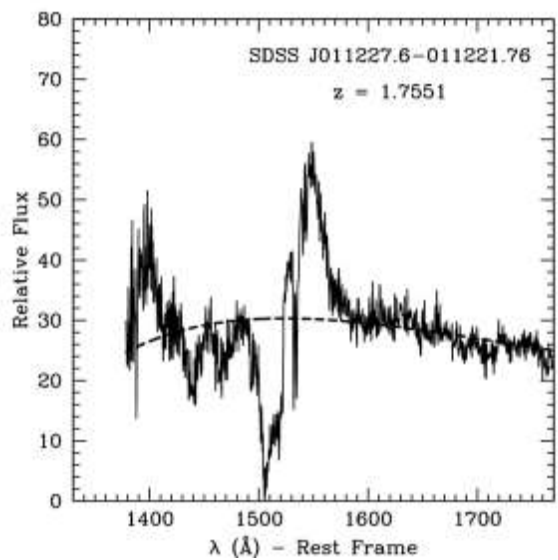

Fig. 4

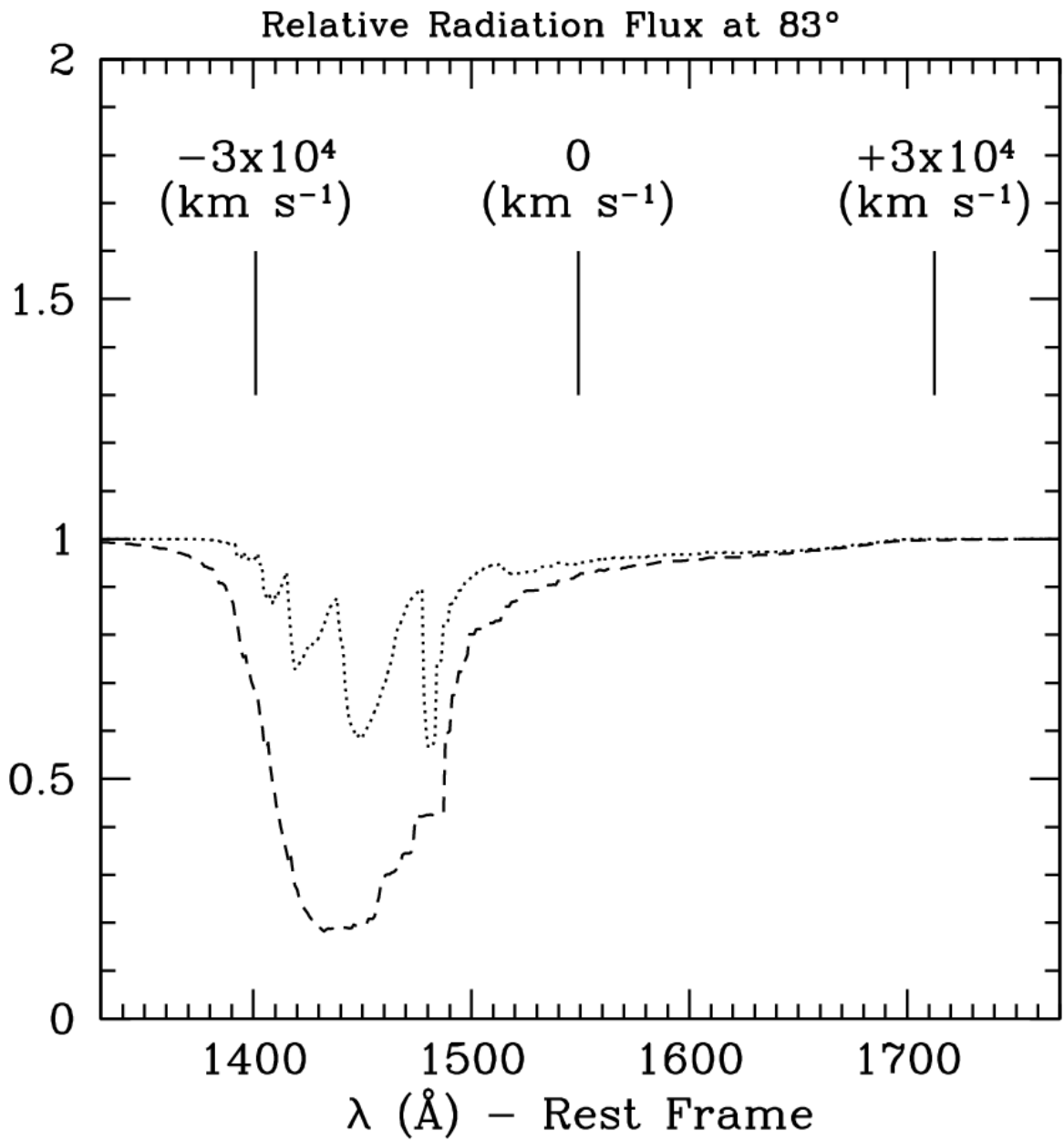

Fig. 5